\input harvmac


\lref\balasubramanian{V. Balasubramanian, J. de Boer, B. Feng, Y.
H. He, M. x. Huang, V. Jejjala and A. Naqvi, ``Multi-trace
superpotentials vs. matrix models,'' Commun. Math. Phys. {\bf
242,} 361 (2003) [arXiv:hep-th/0212082].}

\lref\CachazoRY{ F.~Cachazo, M.~R.~Douglas, N.~Seiberg and
E.~Witten, ``Chiral rings and anomalies in supersymmetric gauge
theory,'' [arXiv:hep-th/0211170].}

\lref\Svrcek{P. Svrcek, ``Chiral Rings, Vacua and Gaugino
Condensation of Supersymmetric Gauge Theories,"
[arXiv:hep-th/0308037].}

\lref\CachazoZK{ F.~Cachazo, N.~Seiberg and E.~Witten, ``Phases of
N = 1 supersymmetric gauge theories and matrices,''
[arXiv:hep-th/0301006].}

\lref\Cachazothree{ F. ~Cachazo, N. ~Seiberg and E. ~Witten,
''Chiral Rings and Phases of ~Supersymmetric Gauge Theories,''
[arXiv:hep-th/0303207].}

\lref\CachazoSp{F. Cachazo, ``Notes on supersymmetric $Sp(N)$
theories with an antisymmetric tensor,'' [arXiv:hep-th/0307063].}

\lref\dorey{N. Dorey, T. J. Hollowood, S. P. Kumar and A.
Sinkovics, ``Exact superpotentials from matrix models,'' JHEP {\bf
0211,} 039 (2002) [arXiv:hep-th/0304138].}

\lref\vafasp{K. Intriligator, P. Kraus, A. V. Ryzhov, M. Shigemori
and C. Vafa, ``On Low Rank Classical Groups in String Theory,
Gauge Theory and Matrix Models,'' [arXiv:hep-th/0311181].}

\lref\Argurio{R. Argurio, G. Ferretti and R. Heise, ``An
Introduction to Supersymmetric Gauge Theories and Matrix Models,''
[arXiv:hep-th/0311066].}

\lref\migdal{A. A. Migdal, ``Loop Equations and $1/N$ Expansion,"
Phys. Rept. {\bf 102}, 199 (1983).}

\lref\morozov{A. Morozov, ``Integrability and Matrix Models,''
Phys.Usp. {\bf 37}, 1 (1994) (No.1) [arXiv:hep-th/9303139].}

\lref\konishig{ K. Konishi and A. Ricco, '' Calculating Gluino
Condensates in ${\cal N}=1$ SYM from Seiberg-Witten Curves,"
[arXiv:hep-th/0306128].}

\lref\merlatti{P. Merlatti, '' Gaugino condensate and Phases of
${\cal N}=1$ Super Yang-Mills Theories," [arXiv:hep-th/0307115].}

\lref\ferraria{F. Ferrari, '' On Exact Superpotentials in
Confining vacua," Nucl. Phys B {\bf 648} (2003) 161
arXiv:hep-th/0210135.}

\lref\ferrarib{F. Ferrari, '' Quantum Parameter Space and Double
Scaling Limits in ${\cal N}=1$ Super Yang-Mills Theory," Phys.
Rev. D {\bf 67} (2003) 085013 arXiv:hep-th/0211069.}

\lref\wz{J. Wess and B. Zumino, ``Consequences of Anomalous Ward
Identities," Phys. Lett. {\bf 37B}, 95 (1971).}

\lref\ferraric{F. Ferrari, '' Quantum Parameter Space in Super
Yang-Mills. II," Phys. Lett. B {\bf 557} (2003) 290
arXiv:hep-th/0301157.}

\lref\monopolecond{ J. de Boer and Y. Oz, ``Monopole Condensation
and Confining Phase of ${\cal N}=1$ Gauge Theories Via M Theory
Fivebrane,'' arXiv:hep-th/9708044.}

\lref\ringsSO{ E.~Witten, ``Chiral Ring Of Sp(N) and SO(N)
Supersymmetric
 Gauge Theory In Four Dimensions.'' arXiv:hep-th/0302194}

\lref\SeibergRS{ N.~Seiberg and E.~Witten, ``Electric - magnetic
duality, monopole condensation, and confinement in N=2
supersymmetric Yang-Mills theory,'' Nucl.\ Phys.\ B {\bf 426}, 19
(1994) [Erratum-ibid.\ B {\bf 430}, 485 (1994)]
[arXiv:hep-th/9407087].}

\lref\amatietal{D. Amati, G. C. Rossi, and G. Veneziano,
``Instanton Effects In Supersymmetric Gauge Theories,'' Nucl.
Phys. {\bf B249} (1985) 1.}

\lref\dhkm{N. Dorey, T. J. Hollowood, V. Khoze, and M. P. Mattis,
``The Calculus Of Many Instantons,'' hep-th/0206063, Phys. Rept.
{\bf 371} (2002) 231.}

\lref\russians{V. A. Novikov, M. A. Shifman, A. I. Vainshtein, M.
B. Voloshin, and V. I. Zakharov, ``Supersymmetry Transformations
Of Instantons,'' Nucl. Phys. {\bf B229} (1983) 394; V. A. Novikov,
M. A. Shifman, A. I. Vainshtein, and V. I. Zakharov, ``Exact
Gell-Mann-Low Functions of Supersymmetric Yang-Mills Theories From
Instanton Calculus,'' ``Instanton Effects In Supersymmetric
Theories,'' Nucl. Phys. {\bf B229} (1983) 381, 407.}

\lref\ArgyresJJ{ P.~C.~Argyres and M.~R.~Douglas, ``New phenomena
in SU(3) supersymmetric gauge theory,'' Nucl.\ Phys.\ B {\bf 448},
93 (1995) [arXiv:hep-th/9505062].}

\lref\soone{C. Ahn and Y. Ookouchi, ``Phases of ${\cal N}=1$
Supersymmetric SO/Sp Gauge Theories via Matrix Model,''
[arXiv:hep-th/0302150].}

\lref\sotwo{L. F. Alday and M. Cirafici, ``Effective
Superpotentials via Konishi Anomaly,'' [arXiv:hep-th/0204119]. }

\lref\krausone{P. Kraus and M. Shigemori, `` On the matter of the
Dijkgraaf-Vafa conjecture,'' JHEP {\bf 0304,} 052 (2003)
[arXiv:hep-th/0304138].}

\lref\kraus{P. Kraus, A. V. Ryzhov and M. Shigemori, ``Loop
equations, matrix models, and ${\cal N}=1$ supersymmetric gauge
theories,'' [arXiv:hep-th/0304138].}

\lref\hungerford{ T. W. Hungerford, {\it Algebra }
(Springer-Verlag, 1974).}

\lref\konishione{ K.~Konishi, ``Anomalous Supersymmetry
Transformation Of Some Composite Operators In Sqcd,'' Phys.\
Lett.\ B {\bf 135}, 439 (1984).}

\lref\konishitwo{ K.~i.~Konishi and K.~i.~Shizuya, ``Functional
Integral Approach To Chiral Anomalies In Supersymmetric Gauge
Theories,'' Nuovo Cim.\ A {\bf 90}, 111 (1985).}

\lref\gtransitions{ F.~Cachazo, K.~A.~Intriligator and C.~Vafa,
``A large N duality via a geometric transition,'' Nucl.\ Phys.\ B
{\bf 603}, 3 (2001) [arXiv:hep-th/0103067].}

\lref\vafawarner{C.~Vafa and N.P. Warner, ``Catastrophes and the
classification of conformal theories,'' Phys. Lett. B {\bf 218},
51 (1989) ; W. Lerche, C. Vafa and N. P. Warner, ``Chiral Rings in
N=2 Superconformal Theories,'' Nucl. Phys. B {\bf 324}, 437
F(1989).}

\lref\grassmannian{ E.~Witten, ``The Verlinde Algebra And The
Cohomology Of The Grassmannian,'' arXiv:hep-th/9312104, and in
{\it Quantum Fields And Strings: A Course For Mathematicians}, ed.
P. Deligne et. al. (American Mathematical Society, 1999), vol. 2,
pp. 1338-9.
}

\lref\SeibergPQ{ N.~Seiberg, ``Electric - magnetic duality in
supersymmetric nonAbelian gauge theories,'' Nucl.\ Phys.\ B {\bf
435}, 129 (1995) [arXiv:hep-th/9411149].
}

\lref\IntriligatorID{ K.~A.~Intriligator and N.~Seiberg,
``Duality, monopoles, dyons, confinement and oblique confinement
in supersymmetric SO(N(c)) gauge theories,'' Nucl.\ Phys.\ B {\bf
444}, 125 (1995) [arXiv:hep-th/9503179].
}

\lref\KutasovSS{ D.~Kutasov, A.~Schwimmer and N.~Seiberg, ``Chiral
Rings, Singularity Theory and Electric-Magnetic Duality,'' Nucl.\
Phys.\ B {\bf 459}, 455 (1996) [arXiv:hep-th/9510222].
}

\lref\DijkgraafFC{ R.~Dijkgraaf and C.~Vafa, ``Matrix models,
topological strings, and supersymmetric gauge theories,''
arXiv:hep-th/0206255.
}

\lref\spone{P.L. Cho and P. Kraus, ``Symplectic SUSY gauge
theories with antisymmetric matter,'' Phys. Rev. D {\bf 54,} 7640
(1996) [arXiv:hep-th/9607200].}

\lref\sptwo{C. Csaki, W. Skiba and M. Schmaltz, ``Exact results
and duality for $Sp(2N)$ SUSY gauge theories with an antisymmetric
tensor,'' Nucl. Phys. B {\bf 487,} 128 (1997)
[arXiv:hep-th/9607210].}

\lref\DijkgraafVW{ R.~Dijkgraaf and C.~Vafa, ``On geometry and
matrix models,'' arXiv:hep-th/0207106.
}

\lref\GM{A.Gerasimov, A.Marshakov, A.Mironov, A.Orlov et al.
             Nucl.Phys. {\bf B357} (1991) 565}

\lref\MM{A.Mironov et al. Phys.Lett. {\bf 252B} (1990) 47;
             J.Ambjorn, J.Jurkiewicz and Yu. Makeenko, Phys.Lett.
                 {\bf 251B} (1990) 517;
             H.Itoyama and Y.Matsuo, Phys.Lett. {\bf 255B} (1991)
             202.}

\lref\DijkgraafDH{ R.~Dijkgraaf and C.~Vafa, ``A perturbative
window into non-perturbative physics,'' arXiv:hep-th/0208048.
}
\lref\DouglasNW{ M.~R.~Douglas and S.~H.~Shenker, ``Dynamics of
SU(N) supersymmetric gauge theory,'' Nucl.\ Phys.\ B {\bf 447},
271 (1995) [arXiv:hep-th/9503163].
}

\lref\VafaPR{R. Dijkgraaf, M. T. Grisaru, C. S. Lam, C. Vafa and
D. Zanon, ``Perturbative computation of glueball
superpotentials,''[arXiv:hep-th/0211017].}

\lref\fluxes{F.~Cachazo and C.~Vafa, ``N = 1 and N = 2 geometry
from fluxes,'' arXiv:hep-th/0206017.
}

\newbox\tmpbox\setbox\tmpbox\hbox{\abstractfont }

 \Title{\vbox{\baselineskip12pt \hbox{hep-th/0311238 }
\hbox{PUPT-2101}
 }}
{\vbox{\centerline{On Nonperturbative Exactness of Konishi
Anomaly}
\medskip
\centerline{ and}
\medskip
\centerline{the Dijkgraaf-Vafa Conjecture}}}

\smallskip
\centerline{ Peter Svrcek}
\smallskip
\bigskip
\centerline{\it  Joseph Henry Laboratories, Princeton University}
\centerline{\it Princeton, New Jersey 08544, USA}
\bigskip
\vskip 1cm \noindent In this paper we study the nonperturbative
corrections  to the generalized Konishi anomaly that come from the
strong coupling  dynamics of the gauge theory. We consider $U(N)$
gauge theory with adjoint and $Sp(N)$ or $SO(N)$ gauge theory with
symmetric or antisymmetric tensor. We study the algebra of chiral
rotations of the matter field and show that it does not receive
nonperturbative corrections. The algebra implies Wess-Zumino
consistency conditions  for the generalized Konishi anomaly which
are used to show that the anomaly does not receive nonperturbative
corrections  for superpotentials of degree less than  $2l+1$ where
$2l=3c(Adj)-c(R)$ is the one-loop beta function coefficient. The
superpotentials of higher degree can be nonperturbatively
renormalized because of the ambiguities in the UV completion of
the gauge theory. We discuss the implications for the
Dijkgraaf-Vafa conjecture.

\Date{November 2003}

\noindent

\newsec{Introduction}

Recently there has been a renewed interest in the study of ${\cal
N}=1$ supersymmetric gauge theories thanks to the work of
Dijkgraaf and Vafa \DijkgraafFC,\ \DijkgraafVW,\ \DijkgraafDH\ who
used string theory arguments to calculate holomorphic data of  a
large class of gauge theories in terms of an auxiliary matrix
model. The bosonic potential of the matrix model is the
superpotential of the gauge theory. Identifying the generating
function for the glueball moments with the matrix model resolvent,
the effective superpotential of the gauge theory with the massive
scalar  integrated out was related to the planar matrix model free
energy.  For the $U(N)$ gauge theory, the nonperturbative part of
the superpotential comes from the measure of the matrix model and
is given by a sum of Veneziano-Yankielowicz superpotentials of the
$U(N_i)$ subgroups. The perturbative part is given by a sum of
planar diagrams of the matrix model.

The conjecture can be studied without  recourse to string theory
arguments. For a pedagogical introduction to the gauge theory
methods used to study the Dijkgraaf-Vafa conjecture, see
\Argurio.\ The authors of \VafaPR\ gave a field theory argument
showing that the Feynman diagrams contributing to the perturbative
part of the glueball superpotential reduce to matrix model
diagrams. A different approach was pursued  in \CachazoRY\ using
the chiral ring of the gauge theory. The generalized Konishi
anomalies of the chiral rotations of the adjoint field imply
constraints between chiral operators. These constraints have the
same form as the loop equations of the matrix model in the planar
limit. Hence the effective superpotential can be expressed in
terms of the matrix model free energy up to a coupling independent
term which can by seen to be a sum of Veneziano-Yankielowicz
superpotentials by taking the limit of large couplings of the
superpotential.

To complete the above argument it is necessary to verify that the
generalized Konishi anomaly equations remain valid
nonperturbatively and that the low energy effective description of
the gauge theory in terms of the glueball fields is correct. In
\CachazoRY\ it was suggested that one can prove the absence of
corrections to the generalized Konishi anomaly by showing that the
algebra of chiral rotations of the matter field does not have
nonperturbative corrections and then arguing along the lines of
Wess-Zumino consistency conditions that the anomalies do not have
nonperturbative corrections. In this paper we carry out this
proposal. We show that the Konishi anomaly does not have
nonperturbative corrections for superpotentials of degree less
than $2l+1$ where $2l=3c(Adj)-c(R)$ is  the one-loop beta function
coefficient.

The consistency conditions do not completely fix the
nonperturbative corrections to anomaly for superpotentials of a
degree higher than $2l.$ Such corrections are expected due to
ambiguities in the definition of highly nonrenormalizable
operators like $\Tr \, \Phi^n$ \vafasp,\ \dorey\ and
\balasubramanian.\ We show that all the  ambiguities can be
absorbed into nonperturbative redefinition of the superpotential.
There are additional UV ambiguities for gauge theories which are
not asymptotically free coming the freedom in their UV completion.
For these theories our proof does not apply because $\Lambda^{2l}$
has zero or negative dimension, whence there are infinitely many
types of corrections of a given dimension. The consistency
conditions are not powerful enough to constrain these corrections
uniquely. In summary, in this paper we prove the absence of
nonperturbative corrections to the generalized Konishi anomaly
that come from  strong coupling dynamics and determine the form of
corrections  for high degree superpotentials.

The proof can be applied to gauge theories whose algebra of chiral
rotations of matter fields forms an extension of a partial
Virasoro algebra. For example it is possible to consider matter in
other than adjoint representation. In particular we show
nonrenormalization of the generalized Konishi anomaly for $SO(N)$
and $Sp(N)$ gauge theories with matter in the symmetric or
antisymmetric representation. The nonrenormalization of the
generalized Konishi anomaly for $Sp(N)$ with the antisymmetric
tensor is expected in the light of  recent results  \CachazoSp\
and \vafasp\ that demonstrated agreement between the effective
superpotential obtained using Konishi anomalies with the
dynamically generated superpotential approach \spone, \sptwo.\ The
papers \CachazoSp\ and \vafasp\ resolved a puzzle raised in
\krausone, \kraus\ and \sotwo\ about the application of
Dijkgraaf-Vafa correspondence for $Sp(N)$ with antisymmetric
matter.

\bigskip
\noindent {\it Organization and Results of the Paper}

In section 2 we introduce the algebra of chiral rotations of the
matter field and show that it is an ${\cal N}=1$ extension of a
partial Virasoro algebra. We consider the $U(N)$ gauge theory with
adjoint scalar to keep the discussion concrete. In section 3 we
discuss the generalized Konishi anomalies of the chiral rotations
and use the Virasoro symmetry to derive Wess-Zumino consistency
conditions for the anomalies. In section 4 we use $U(1)$
symmetries of the gauge theory to determine the form of the
nonperturbative corrections. In section 5 we use the Lie algebraic
structure of the algebra of chiral rotations to prove that the
algebra cannot get deformed nonperturbatively. This implies that
the Wess-Zumino consistency conditions derived in section 4 are
exact nonperturbatively. We use them to show for $U(N)$ in section
6 and for $SO(N)$ and $Sp(N)$ in section 7 that the generalized
Konishi anomaly cannot have nonperturbative corrections except for
nonperturbative renormalization of superpotentials of degree
greater than $2l=3c(Adj)-c(Matter).$ In section 8 we review the
loop equations of the planar matrix model, considering them as
anomalies of the matrix model free energy under reparametrization
of the matrix $M$  to highlight their similarity with gauge theory
anomalies. In section 9 we discuss the implications of the results
for the Dijkgraaf-Vafa conjecture.

\newsec{The Algebra of Chiral Rotations}

In \CachazoRY,\ a series of constraints for the chiral operators
were derived  by considering the possible anomalies of the chiral
rotations  $\delta \Phi =f(\Phi, W_\alpha)$ of the adjoint scalar
field. These chiral rotations are generated by the operators
\eqn\lnvar{\eqalign{ L_n&= \Phi^{n+1} {\delta \over \delta \Phi},
\cr Q_{n, \alpha}&= {1 \over 4 \pi} W_\alpha \Phi^{n+1} {\delta
\over \delta \Phi} , \cr R_n&= -{1\over 32 \pi^2} W_\alpha
W^\alpha \Phi^{n+1} {\delta \over \delta \Phi}.}}
 The action of the operators \lnvar\  on the single trace chiral operators $u_k=\Tr \, \Phi^k , w_{k, \alpha}={1\over 4\pi} \Tr \, W_\alpha \Phi^k $ and $ r_k= -{1\over 32 \pi^2}\Tr \, W_\alpha^2 \Phi^k $ is  \eqn\lnact{\eqalign{L_n u_k&= k u_{k+n}, \cr Q_{n,\alpha} u_k&= k w_{k+n,\alpha}, \cr & \dots.}} The classical commutation relations of the generators follow from the definitions \lnvar\  \eqn\vira{\eqalign{ [L_m,L_n]&=(n-m)L_{m+n}, \cr [L_m, Q_{n,\alpha}] &=(n-m)Q_{n+m,\alpha}, \cr [L_m, R_n]&= (n-m)R_{m+n}, \cr
\{Q_{m,\alpha},Q_{n,\beta} \}&=- \epsilon_{\alpha \beta}(n-m)
R_{n+m} , \cr [Q_{m, \alpha}, R_n]&=0 , \cr [R_m, R_n]&=0,}} where
$m,n \geq -1.$ The last two commutators are trivially zero in the
chiral ring because the third and higher powers of $W_\alpha$ are
chiral ring descendants. The $L_n$'s form a partial Virasoro
subalgebra which is extended by $Q_{n,\alpha}$'s and $R_n$'s into
a partial ${\cal N}=1$ super-Virasoro algebra.

The scalar $\Phi$ and the  gauge field are in the adjoint
representation of the $U(N)$ gauge group so they do not couple to
the diagonal $U(1)$ gauge field. Hence shifting $W_\alpha$ by an
anticommuting number is a symmetry of the full gauge theory
\CachazoRY.\ If we define the field ${\tilde W}_\alpha =
W_\alpha+4\pi \psi_\alpha$ where $\psi_\alpha$ is an anticommuting
c-number spinor then the generator of the shift symmetry is
$\partial / \partial \psi_\alpha.$ Hence all expressions are
independent of $\psi_\alpha$ when expressed in terms of ${\tilde
W}_\alpha$ and $\Phi.$ The shift symmetry combines the single
trace chiral operators into \eqn\rs{{\tilde r}_k=-{1\over 32
\pi^2}\Tr \, {\tilde W}_\alpha^2 \Phi^k= r_k-\psi_\alpha
w^\alpha_k-\half \psi_\alpha \psi^\alpha u_k.} The shift symmetric
generators of the chiral rotations are $L_n$ and
\eqn\shifts{\eqalign{ {\tilde Q}_{n,\alpha}&={1\over 4\pi} {\tilde
W}_\alpha \Phi^{n+1} {\partial \over \partial \Phi}
=Q_{n,\alpha}+\psi_\alpha L_n, \cr {\tilde R}_n&= -{1\over 32
\pi^2} {\tilde W}_\alpha^2 \Phi^{n+1} {\partial \over \partial
\Phi} = R_n -\psi_\alpha Q_n^\alpha -\half \psi_\alpha \psi^\alpha
L_n.}}

Shift invariance implies that the commutation relations can be
written in terms of $L_n, {\tilde Q}_{n,\alpha}$ and ${\tilde
R}_n.$ We find that the shift invariant commutation relations are
\eqn\viras{\eqalign{[L_m, {\tilde R}_n]&=(n-m){\tilde R}_{m+n},
\cr \{{\tilde Q}_{m,\alpha}, {\tilde Q}_{n,\beta}
\}&=-\epsilon_{\alpha,\beta}(n-m) {\tilde R}_{m+n}, \cr [{\tilde
R}_m, {\tilde R}_n]&=0.}} We did not write down the $[L,{\tilde
Q}]$ and $[{\tilde Q},{\tilde R}]$ commutators because they are
contained in the $[L,{\tilde R}]$ and $[{\tilde R},{\tilde R}]$
commutators respectively. For future reference let us show that
the first and the third commutation relation in  \viras\ imply the
remaining relation. The first commutator contains the $[L,L],
[L,Q]$ and $[L,R]$ commutators. If we expand the last commutator
in $\psi_\alpha,$ all commutators are trivially zero except for
the commutator multiplying $\psi_\alpha \psi^\alpha$ which is
\eqn\psicom{[L_m, R_n]+[R_m,L_n]+\epsilon_{\alpha\beta}
\{Q_m^\alpha, Q_n^\beta\}=0.}  We use this equation together with
the $[L,R]$ commutator to get the $\{Q, Q\}$ commutator. Hence,
the first and third commutator in \viras\ contain all commutation
relations of the partial  ${\cal N}=1$ super-Virasoro algebra.

\newsec{Wess-Zumino Consistency Conditions for the Generalized Konishi Anomaly}

Assume that the adjoint scalar has the tree level superpotential
\eqn\superpotential{W(\Phi)= \sum_{i=1}^{n+1} {g_i \over i} \Tr \,
\Phi^{i}.} The effective superpotential of the gauge theory is
\eqn\effsuperpotential{\exp \left(-\int d^2\theta W_{eff}
\right)=\left\langle \exp \left(-\int d^4x d^2\theta W(\Phi)
\right) \right\rangle ,} where the path integral is over the
massive fields in the presence of a slowly varying background
gauge field. The effective superpotential has an anomaly under the
chiral rotations generated by $L_n, Q_{n, \alpha}, R_n$
\eqn\lna{\eqalign{L_n W_{eff}&= {\cal L}_n, \cr Q_{n, \alpha}
W_{eff}&={\cal Q}_{n,\alpha},\cr R_n W_{eff}&= {\cal R}_n.}} The
perturbative anomaly of the effective superpotential under the
chiral rotations ${\tilde R}_n$ were derived in \CachazoRY\
\eqn\ln{{\tilde {\cal R}}_k= \sum_{i=1}^{n+1} g_i \tilde{r}_{k+i}-
\sum_{i=0}^k \tilde{r}_i \tilde{r}_{k-i}.} The equation \ln\ is
obtained from the $1/z^{k+2}$ term of the equation (4.14) of
\CachazoRY\ for the generating function for the generalized
Konishi anomaly remembering that the $g_i$ in this paper is
$g_{i-1}$ of \CachazoRY.\ The first part of ${\tilde {\cal R}}_k$
is the classical variation of the superpotential and the second
part comes from the anomalous transformation of the measure of
$\Phi$ under the chiral rotation. The anomalous divergence of the
currents generating the chiral rotations is the Konishi anomaly
\eqn\acu{\eqalign{{\bar D}_\alpha {\bar D}^\alpha J_{L_n}&= {\cal
L}_n, \cr \dots .}} Hence the generalized Konishi anomaly, being
${\bar D}_\alpha$ exact, is a chiral ring descendant. Setting \ln\
to zero gives nontrivial relations between the chiral operators,
which enabled the authors of \CachazoRY\ to study the dynamics of
the gauge theory and to give a partial proof of the Dijkgraaf-Vafa
conjecture. We will return to this in more detail in the last
section.

The Lie algebra structure of the chiral rotations implies
relations between anomalies of different chiral rotations. These
conditions were first discussed by Wess and Zumino \wz.\ They
express the closure of the Lie algebra under commutation
relations. For two chiral rotations $R_1$ and $R_2$ the anomaly of
the effective superpotential under $R_1 R_2- R_2 R_1$ must be the
same as the anomaly under $R_3=[R_1,R_2]$ \eqn\wzr{R_1 {\cal
R}_2-R_2 {\cal R}_1= {\cal R}_{[R_1,R_2]}.} The Wess-Zumino
consistency conditions  for the algebra of chiral rotations \vira\
are \eqn\wzc{\eqalign{L_m{\cal L}_n-L_n{\cal L}_m&=(n-m){\cal
L}_{n+m}, \cr L_m {\cal Q}_{n,\alpha}-Q_{n,\alpha} {\cal
L}_m&=(n-m) {\cal Q}_{n,\alpha}, \cr &\dots. }} In the shift
invariant notation, we have \eqn\wzcs{\eqalign{L_m{\tilde {\cal
R}}_n-{\tilde R}_n{\cal L}_m&=(n-m){\tilde {\cal R}}_{m+n}, \cr
{\tilde Q}_{m,\alpha}{\tilde {\cal Q}}_{n,\beta}+{\tilde
Q}_{n,\beta}{\tilde {\cal Q}}_{m,\alpha}&=-\epsilon_{\alpha
\beta}(n-m) {\tilde {\cal R}}_{m+n},\cr {\tilde R}_m{\tilde {\cal
R}}_n-{\tilde R}_n {\tilde {\cal R}}_m&=0.}}

Let us verify that the perturbative anomaly \ln\ satisfies the
Wess-Zumino consistency conditions. The calculations are routine
so we will check only the first equation in \wzc.\ Expanding \ln\
with respect to $\psi_\alpha$ we find using \rs\ and \shifts\
\eqn\lna{{\cal L}_k = \sum_{i=1}^{n+1} g_i u_{k+i} -
2\sum_{i=0}^{k} u_i r_{k-i}.} The action of $L_k$ on ${\cal L}_l$
is \eqn\lka{L_k{\cal L}_l=\sum_{i=1}^{n+1} (k+i)g_i
 u_{l+k} -2\sum_{i=0}^l \left(i \tilde u_{i+k}
r_{l-i}+(l-i) u_i r_{k+l-i}\right).} Subtracting  from this the
analogous expression for $L_l{\cal L}_k$ we get
\eqn\comln{L_k{\cal L}_l-L_l{\cal L}_k=(l-k){\cal L}_{k+l}} which
is the Wess-Zumino consistency condition \wzc\ for the Virasoro
subalgebra.

\newsec{Nonpertubative Corrections}

In this section we review the argument for the absence of the
multi-loop corrections to the generalized Konishi anomaly and then
discuss the structure of nonperturbative corrections. For this it
is instrumental to study the $U(1)$ symmetries of the gauge
theory. The gauge theory has two continuous symmetries, a standard
$U(1)_R$ symmetry and a symmetry $U(1)_\Phi$ under which the
entire superfield $\Phi$ undergoes a rotation \eqn\huffo{ \Phi\to
e^{i\alpha}\Phi. } We also introduce a linear combination of
these, $U(1)_\theta$, which is convenient in certain arguments.
These symmetries are symmetries of the theory with nonzero
superpotential if we assign nonzero $U(1)$ charges to the
couplings $g_k.$

 \eqn\uonecharges{ \matrix{ & \Delta & Q_\Phi & Q_R &
Q_\theta \cr
 \Phi & 1 & 1 & 2/3 &  0 \cr
W_\alpha & 3/2 & 0 & 1 & 1 \cr
 g_k & 1-k & -k & {2\over 3} (1-k) & 2 \cr
 \Lambda^{2l}& 2l & 2l & 4l/3 & 0 \cr
{\tilde{\cal R}}_k & 6+k& k &  4+2k/3& 4 \cr}} The one-loop beta
function coefficient is $2l=3c(Adj)-c(R)$ where $c(R)$ is the
index of the representation $R$ of the matter field
\eqn\casimirs{\matrix{R&
U_{Adj}(N)&SO(N)_A&SO(N)_S&Sp(N)_A&Sp(N)_S \cr
c(R)&N&N-2&N+2&N-1&N+1 \cr l& N& N-2&N-3&N+2&N+1}} The shift
invariant ${\tilde W}_\alpha$ and the anticommuting shift c-number
$ \psi_\alpha$ have the same $U(1)$ charges as $W_\alpha.$  These
symmetries are violated at  one loop. In the last line of the
table \uonecharges\ we have written the charges by which the
anomaly ${\tilde {\cal R}}_k$ violates the $U(1)$ symmetries. The
higher loop computations are finite and the $U(1)$ symmetries
leave them invariant.

We are now ready to analyze the corrections to the generalized
Konishi anomaly \ln.\ The corrections must have the same $U(1)$
charges as ${\tilde {\cal R}}_k.$ They are polynomial in the
chiral operators. Furthermore,  the corrections that depend on
$g_k$ must vanish for the theory with zero superpotential and the
nonperturbative corrections that depend on $\Lambda^{2l}$ vanish
when we take the strong coupling scale $\Lambda$ to zero. Hence,
the corrections to the anomaly are also polynomial in $g_k$ and
$\Lambda^{2l}.$ Referring to the table \uonecharges\ we see that
the only polynomials in $g_k,$ $\Phi$ and $W_\alpha$ with the
quantum numbers of ${\tilde {\cal R}}_k$ are the ones already
present in the one loop expression \ln.\ Hence the anomaly does
not have higher loop contributions, as claimed at the end of the
previous paragraph. The  nonperturbative corrections are
polynomial in $\Lambda^{2l}.$ The possible $j$ instanton
corrections to ${\tilde {\cal R}}_k$ are of the form
$\Lambda^{2jl} g_{i+2jl}{\tilde r}_{k+i}$ and $\Lambda^{2jl}
{\tilde r}_{i-2jl} {\tilde r}_{k-i}.$

We can similarly derive  the possible form of corrections to the
extended Virasoro algebra \vira.\ The corrections to the $[L,L]$
commutator are linear the Virasoro generators $L_n$  and
polynomial in $g_k$ and $\Lambda^{2h}.$ The Virasoro generator
$L_n$ \lnvar\ increases the $U(1)$ charges of a chiral operators
by the same value as multiplication by $\Phi^n.$  Hence, the
commutator $[L_m, L_n]$ fixes $Q_\theta$ and increases the
dimension by $m+n.$ Consulting the table \uonecharges\ we see that
$g_i$ has $Q_\theta=2$ charge so the there are no corrections that
depend on the superpotential. The nonperturbative $l$ instanton
corrections have the form $\Lambda^{2jl} L_{m+n-2jl}.$ Similar
corrections contribute to the $[L,Q],[L,R]$ and ${Q,Q}.$ The
commutators that shift $Q_\theta$ by two  can also have
corrections proportional to $g_i.$ Counting the $U(1)$ charges we
see that the $[L_m,R_n]$ commutator has corrections $\Lambda^{2jl}
R_{m+n-2jl}$ and $\Lambda^{2jl} g_i L_{m+n+i-2jl}.$ There are
similar corrections to $\{Q,Q\}.$ The $[Q,R]$ and $[R,R]$
commutators cannot have corrections because they map chiral
operators into chiral ring descendants.

\newsec{Nonrenormalization of the Algebra of Chiral Rotations}

In this section we prove the nonrenormalization of the algebra
\vira\ of chiral rotations of the $U(N)$ adjoint scalar. Firstly
we analyze in detail the corrections to the partial Virasoro
subalgebra \eqn\comln{[L_m, L_n]=(n-m)L_{m+n} +\sum_{j>0}
\Lambda^{2jN}b^j_{m,n}L_{m+n-2jN},} where the coefficients
$b^j_{m,n}$ are antisymmetric in $m$ and $n$ by antisymmetry of
the commutator \comln.\ The coefficient $b^j_{m,n}$ is in front of
$L_{m+n-2jN}$ hence it vanishes if $m+n-2jN<-1$ because $L_{-1}$
is the lowest nonzero generator. We will prove that all the
coefficients $b^j_{m,n}$ can be absorbed into nonperturbative
redefinition of the Virasoro generators \eqn\virdef{{\rm
L}_n=L_n+\sum_{j>0}a^j_n \Lambda^{2jN}L_{n-2jN},} where $a^j_n$ is
the coefficient of the $j$-instanton correction to ${\rm L}_n.$
The Virasoro generators are corrected which is natural considering
that they act on the nonperturbatively corrected chiral operators
${\tilde r}_k.$ In terms of the new basis of generators  $ {\rm
L}_n$ the commutations relations of the partial Virasoro algebra
remain valid nonperturbatively \eqn\virasoro{[{\rm L}_m,{\rm
L}_n]=(n-m){\rm L}_{m+n}.}

Calculating the coefficients $a^j_n$ which parameterize the
nonperturbative corrections to ${\rm L}_n$'s is beyond the scope
of the present paper. We will show instead that there is a choice
of $a^j_n$'s for which the Virasoro algebra takes the standard
form \virasoro.\ This shows that the algebra itself is not
corrected even though the Virasoro operators might receive
corrections. We make induction in the instanton number of the
nonperturbative corrections.  The coefficients $b^j_{m,n}$ obey
equations that follow from the Jacobi identity
\eqn\Jacobi{[L_l,[L_m,L_n]]+[L_n,[L_l,L_m]]+[L_m,[L_n,L_l]]=0.} On
the zero instanton level the identity reduces to the Jacobi
identity for the  Virasoro algebra which is satisfied. On the one
instanton level, we evaluate the commutators in \Jacobi\ using
\comln\ to find the coefficient of the $\Lambda^{2N}L_{l+m+n-2N}$
term which has to be
zero\eqn\oneb{(n-m)b^1_{l,m+n}+(m+n-l-2N)b^1_{m,n}+ cyclic \,
\,permutations=0.} The one instanton corrections can be absorbed
into one instanton corrections to $L_n$'s \virdef.\ The new
commutation relations are \eqn\rcom{[{\rm L}_m,{\rm
L}_n]=(n-m){\rm L}_{m+n}+{\rm b}^1_{m,n} \Lambda^{2N} {\rm
L}_{m+n-2N}+ \dots,} where ${\rm b^1_{m,n}}$'s are the redefined
nonperturbative corrections \eqn\onein{{\rm
b}^1_{m,n}=b^1_{m,n}+(n-m-2N)a^1_n+ (n-m+2N)a^1_m-(n-m)a^1_{m+n}.}
We show  that $b^1_{m,n}$ can be set to zero by redefinition ${\rm
L}_{m+n}=L_{m+n}+a^1_{m+n}\Lambda^{2N} {\rm L}_{m+n-2N}$ by
induction on $m+n.$ The first step of the induction holds because
$b^1_{m,n}$ vanishes for $m+n<2N-1.$ By induction hypothesis we
assume that we have redefined $L_{m+n}$ for $m+n<M$ so that
$b^1_{m,n}=0.$  Setting $l,m,n$ in equation \oneb\ equal to
$0,m,M-m$ respectively, we find for $0<m<M$
\eqn\onc{(M-2m)b^1_{0,M}+(M-2N)b^1_{m,M-m}+ (m-M)b^1_{m,M-m}+m
b^1_{M-m,m}=0 .} Using antisymmetry of $b^1_{m,n}$ in $m$ and $n$
we rewrite this as \eqn\rest{2N b^1_{m,M-m}=(M-2m)b^1_{0,M}.} From
\onein\ the redefined nonperturbative corrections  are
\eqn\bn{\eqalign{{\rm b}^1_{0,M}&=b^1_{0,M}-2N a^1_M ,\cr {\rm
b}^1_{m,M-m}&=b^1_{m,M-m}-(M-2m)a^1_M.}} We see from \rest\ that
taking $a_M=b^1_{0,M}/2N$ we  set ${\rm b}^1_{m,n}=0$ for $m+n=M.$
This completes the induction in $m+n$ and shows that there are no
one instanton corrections to the Virasoro algebra. We can now
proceed with the induction in the instanton number by assuming
absence of nonperturbative corrections to the Virasoro algebra for
instanton number less than $k.$ We also assume that we have
redefined the the Virasoro operators ${\rm L}_n$ up to instanton
number $k-1$ to set $b^j_{m,n}=0$  for $j<k.$ The proof that the
$k$ instanton corrections to the Virasoro algebra can be absorbed
into $k$ instanton redefinition of the operators ${\rm L}_n$ goes
exactly as the above calculation in the one instanton case because
the necessary equations at the $\Lambda^{2kN}$ order are identical
to the equations $\oneb,\onein-\bn $ we found at $\Lambda^{2N}$
order after substituting  $N$ for $kN$ in all equations.  The
additional terms in \oneb\ and \onein\ that would come from lower
instanton corrections  vanish by the induction hypothesis.

Now it remains to show that the commutation relations of ${\rm
L}_{-1}=L_{-1}$ with ${\rm L}_n$ do not get corrected. Firstly
consider one instanton corrections. Notice that $b^1_{-1,0}$
vanishes on dimensional grounds as noted below \comln.\ Taking
$l,m,n$ in \oneb\ to be $-1,0,n$ for $n> 0$ we find
$2Nb^1_{1,n}=0$ which completes the proof of the absence of one
instanton corrections. We prove the absence of $k$ instantons
corrections the same way after substituting $N$ for $kN$ in
\oneb.\

We give two different proofs of the nonrenormalization of the
remaining commutators of the algebra of chiral rotations. The
first one is simpler and uses the shift symmetry of the
commutations relations. The second one does not use the $U(1)$
shift symmetry and hence is applicable for the $SO(N)$ and $Sp(N)$
gauge theories as well. We postpone it to the Appendix A because
it is more technical. From now on we do not use roman font to
distinguish the nonperturbatively defined generators.

Let us outline the first argument. We use shift symmetry to fix
the nonperturbative definitions ${\rm Q}_{n,\alpha}, {\rm R}_n$
for $n\geq 2N$ using the nonperturbatively defined $L_n$ \virdef.\
The last commutator in \viras\  \eqn\virar{[{\tilde R}_m, {\tilde
R}_n]=0} cannot receive nonperturbative corrections. Its lowest
$\psi_\alpha $ component is the $[{\rm R},{\rm R}]=0$ commutator
which has to vanish in the chiral ring because the commutator
shifts $\Phi$ by a chiral operator containing the fourth power of
$W_\alpha.$ But the third and higher powers of $W_\alpha$ are
chiral ring descendants, so the commutator has trivial action in
the chiral ring. The nonperturbative corrections to the first
commutator in \viras\ that are allowed by shift symmetry  are
\eqn\virasf{[L_m,{\tilde R}_n]=(n-m){\tilde R}_{m+n} +
\sum_{j=1}^\infty \sum_{i=1}^{n+1}\Lambda^{2jN} g_i c^{i,j}_{m,n}
L_{m+n+i-2jN}} because the  $\psi_\alpha^2$ component of \virasf\
is the $[L,L]$ commutator, which does not have nonperturbative
corrections. The nonperturbative corrections \virasf\ contribute
to the $[L,R]$ commutator only. To prove that these corrections
vanish we evaluate the $L,Q,R$ Jacobi identity
\eqn\lqrjacobi{\eqalign{[Q_{l,\alpha},[L_m,R_n]]+[L_m,[R_n,Q_{l,\alpha}]]+[R_n,[Q_{l,\alpha},L_m]]&=\cr
=[Q_{l,\alpha}, \sum_{j=1}^{\infty} \sum_{i=1}^{n+1}
\Lambda^{2jN}g_i c^{i,j}_{m,n} L_{m+n+i-2jN}]&=\cr
=\sum_{j=1}^{\infty} \sum_{i=1}^{n+1}\Lambda^{2jN}g_i
(m+n-l+i-2jN) c^{i,j}_{m,n} Q_{m+n+i+l-2jN,\alpha}&=0.}} In
simplifying \lqrjacobi\ we used the $[L,Q]$ commutator \virasf\
which is nonrenormalized  by shift symmetry and  the $[R,Q]=0$
commutator. Clearly, the only way to satisfy the Jacobi identity
\lqrjacobi\ is that $c^{i,j}_{m,n}=0.$ All corrections to \virasf\
vanish. Hence, none of the commutation relations of the extended
Virasoro algebra get nonperturbative corrections because as we
noted below \viras\ the above two commutators imply the remaining
one.

\newsec{ Nonperturbative Corrections to the Konishi Anomaly for $U(N)$ Gauge Theory}

Let us now consider nonperturbative corrections to the anomaly.
The anomaly ${\tilde {\cal R}} _k$ \ln\ differs from its
perturbative value implicitly through the dependence of the chiral
operators ${\tilde r}_k$ on nonperturbative physics. In this
section we ask the question whether there are additional
nonperturbative corrections that depend explicitly on
$\Lambda^{2jN}.$ We can easily introduce terms proportional to
$\Lambda^{2jN}$ into the expression for ${\cal R}_k$ by redefining
the chiral operators \eqn\red{{\tilde r}_k={\tilde r}_k+\alpha
\Lambda^{2N} {\tilde r}_{k-2N}+ \dots.} Notice that $r_k$ for
$k>1$ are nonrenormalizable operators so their value depends on
the renormalization scheme. It is natural to expect terms of the
form \red\ to relate the definitions of $r_k$ coming from
different renormalization schemes. Hence we expect that the
anomaly has generically terms proportional to $\Lambda^{2jN}$ if
we take some arbitrary prescription for ${\tilde r}_k.$

However, there is a natural definition of the higher moments
${\tilde r}_k.$ In the previous section we showed that there is a
preferred basis for the generators of the chiral rotations
${\tilde  R}_k$ in terms of which  the partial super-Virasoro
algebra takes the standard form \viras.\ We can use their action
on the chiral operators to give a nonperturbative definition of
nonrenormalizable operators ${\tilde r}_k$ for $k>1$ in terms of
the the first moment ${\tilde r}_k = { L}_k {\tilde r}_1.$ It
follows from the commutation relations \viras\ that remaining
operators ${\tilde R}_k$ act on the chiral operators as before
\lnact.\

Having defined ${\tilde r}_k$ nonperturbatively, we can now show
using the Wess-Zumino consistency conditions that the one-loop
anomaly $\sum {\tilde r}_i{\tilde r}_{k-i}$ in the path integral
measure for $\Phi$ does not have nonperturbative corrections. We
will also show that the consistency conditions allow
nonperturbative renormalization of the superpotential. The
consistency conditions of the full gauge theory \wzcs\ do not have
nonperturbative corrections because their derivation rested only
on the commutation relations of the super-Virasoro algebra \viras\
which are nonrenormalized. We deduced in section 4 using  $U(1)$
symmetries that the general form of nonperturbative corrections to
${\tilde {\cal R}}_n$ is \eqn\lkc{\eqalign{{\tilde {\cal R}}_k
=&\sum_i (g_i +\Lambda^{2N}g_{i+2N}
c^1_{k,i}+\dots)\tilde{r}_{k+i}+ \cr &-\sum_{i=0}^k \tilde{r}_i
\tilde{r}_{k-i} - \Lambda^{2N} \sum_{i=2N}^{k} d^1_{k,i}
\tilde{r}_{i-2N} \tilde{r}_{k-i}+ \dots.}} In writing \lkc\ we
take $g_k=0$ for $k<1$ and $k>n+1$ to simplify the notation. We
can consider the corrections to the the superpotential separately
from the corrections to the one-loop anomaly. The corrections to
the superpotential are proportional to $\Lambda^{2jN}g_{i+2jN}$
which have the same quantum numbers as $\tilde{r}_{-i}$ which does
not exist. Hence the two types of corrections do not mix.

Firstly we show that all the nonperturbative corrections  to the
one-loop part of ${\tilde {\cal R}}_k$ vanish. Notice, that the
lowest dimensional correction is ${\tilde r}_0 {\tilde r}_0
\Lambda^{2N}$  which contributes to ${\tilde {\cal R}}_{2N},$
hence the one-loop parts ${\tilde {\cal R}}_k$ for
$k=-1,0,\dots,2N-1$ does not have nonperturbative corrections. The
first consistency condition \wzcs\ with $m=0$ simplifies to $L_0
{\tilde {\cal R}}_k= k {\tilde {\cal R}}_k$ because $R_k {\tilde
r}_0 {\tilde r}_0=0.$ In other words $L_0$ acting on ${\tilde
{\cal R}}_k$ gives $k$ times the anomaly. But $L_0$ acting on  a
$j$-instanton correction $\Lambda^{2jN}{\tilde r}_{i-2jN}{\tilde
r}_{k-i}$ gives back $k-2jN$ multiple of the correction, whence
all nonperturbative corrections to the one-loop part of the
anomaly vanish.

It remains to consider the corrections to the classical part of
${\tilde {\cal R}}_n.$ We find from \lkc\ that the first
consistency condition \wzcs\ becomes \eqn\lncomm{\eqalign{{
L}_k{\tilde {\cal R}}_l-{\tilde {R}}_l{\cal L}_k=&(l-k){\tilde
{\cal R}}_{l+k}\cr &+
\sum_{j\geq1}\Lambda^{2jN}\sum_{i=-2jN}^{n-2jN}
[(l+1)c^j_{l,i}-(k+1)c^j_{k,i}-(l-k)c^j_{k+l,i}]g_{i+2jN}
\tilde{r}_{k+l}.}} But the Wess-Zumino consistency conditions do
not have nonperturbative corrections whence we set the terms in
the square brackets to zero
\eqn\imp{(l+1)c^j_{l,i}-(k+1)c^j_{k,i}=(l-k)c^j_{k+l,i}.} Taking
$l=0$ we have $c^j_{k,i}=c^j_{0,i}.$ Clearly, this solves all the
constraints coming from \imp.\ Notice that the terms
$\Lambda^{2jN}c^j_{-1,i} {\tilde r}_{i-1}$ in ${\tilde {\cal
R}}_{-1}$ are absent for $i<1$ because ${\tilde r}_k \sim \Tr
{\tilde W}^2 \Phi^k$ is defined only for positive $k.$ Hence
$c^j_{k,i}=0$ for $i<1.$ In conclusion, the general form of the
anomaly is \eqn\anc{{\tilde {\cal R}}_k =\sum_{i=1}^{n+1} {\rm
g}_i\tilde{r}_{k+i}-\sum_{i=0}^k \tilde{r}_i \tilde{r}_{k-i}}
where \eqn\newg{{\rm g}_i=g_i+\Lambda^{2N}c^1_{0,i} g_{i+2N}+
\Lambda^{4N}c^2_{0,i} g_{i+4N} + \dots.} are the nonperturbatively
renormalized coefficients of the superpotential. Hence, all
corrections to the classical part of the anomaly allowed by the
Wess-Zumino consistency conditions can be absorbed into
nonperturbative renormalization of the superpotential
\eqn\rensup{{\ W}(\Phi)=\sum_{i=1}^{n+1} {{\rm g}_i \over i} \Tr
\, \Phi^{i}.} The  superpotentials of degree less than  $2N+1$
cannot have noperturbative corrections. This is the only ambiguity
that is not fixed by the consistency conditions. We could have
anticipated it from the observation that both $g_i$ and
$\Lambda^{2N}$ are invariant under the chiral rotations hence
substituting for $g_i$ any polynomial $g_i(g_k, \Lambda^{2N})$
with the correct quantum numbers cannot spoil the Wess-Zumino
consistency conditions whose validity depends only on the Lie
algebraic structure of the chiral rotations. As noted around \red\
the nonperturbative corrections depend on the scheme used to
define the single trace operators ${\tilde r}_k.$ Using a
different UV completion of the gauge theory changes the definition
of the chiral operators hence it redefines the superpotential. For
further discussion of Dijkgraaf-Vafa conjecture for high degree
superpotentials, see \vafasp\ \dorey\ and \balasubramanian.\

\newsec{$SO(N)$ and $Sp(N)$ Gauge Theories with Symmetric or Antisymmetric Matter}

In this section we show that the previous analysis applies with
minor modifications to the $SO(N)$ and $Sp(N)$ gauge theories. It
follows that the generalized Konishi anomaly in these gauge
theories does not have nonperturbative corrections for
superpotentials of degree less than  $2l+1.$ Superpotentials of
higher degree might get nonperturbatively renormalized.

The gauge group do not have a decoupled diagonal $U(1)$ subgroup
hence based on the shift symmetry do not carry over from the
$U(N)$ case. That is the main reason why we gave a separate proof
of the nonrenormalizability of the extended Virasoro algebra which
did not use shift symmetry. For simplicity, we do not consider the
fermionic generators and chiral operators. The $SO(N)$ adjoint can
be represented by an  $N \times N$ antisymmetric matrix $\Phi^{\rm
T}= -\Phi.$ The gauge field transforms in the adjoint
representation hence it is antisymmetric as well $W_{\alpha}^{\rm
T}=-W_{\alpha}.$ The $Sp(N)$  has adjoint which can be represented
as $2N \times 2N$ matrix that satisfies the condition $\Phi^{\rm
T}= - J \Phi J^{-1}$ where $J$ is the invariant antisymmetric
tensor of $Sp(N).$ A matrix in the  adjoint representation of
$Sp(N)$ can be written as a product of a symmetric matrix $S$ and
the invariant tensor $\Phi= SJ,$ which explains why this
representation is called symmetric in the literature. The single
trace chiral operators for both gauge groups are $u_{2k}$ and
$r_{2k}$ because the remaining chiral operators vanish by
antisymmetry. Hence the odd coefficients  of the superpotential
\superpotential\  vanish $g_{2k+1}=0.$ Similarly the nonvanishing
generators of the algebra of chiral rotations are $L_{2k}$ and
$R_{2k}$ which form a closed subalgebra of the partial ${\cal
N}=1$ super-Virasoro algebra  \vira.\ Our method also applies to
the symmetric tensor $\Phi^{\rm T}=\Phi$ of $SO(N)$ and the
antisymmetric tensor $\Phi^{\rm T}= J \Phi J^{-1}$ of $Sp(N).$ The
definitions of the representations do not restrict the chiral
operators nor the chiral rotations.

The generalized Konishi anomaly for the $SO(N)$ and $Sp(N)$ gauge
theories has been derived in \soone,\ \sotwo\ and \kraus\
\eqn\sospa{\eqalign{{\cal L}_{k}&=\sum_{i=1}^{n+1} g_{i} u_{i+k}
-\sum_{i=0}^{k} u_{i} r_{k-i} + c_k(R) r_{k}, \cr {\cal R}_{k}&=
\sum_{i=1}^{n+1} g_{i} r_{i+k} - \half \sum_{i=0}^{k} r_{i}
r_{k-i}}} where $c_k(R)$ depends on the representation $R$ of the
matter field  \eqn\cks{\matrix{R&SO_A(N)& SO_S(N)&
Sp_A(N)&Sp_S(N)\cr c_k(R) & 2&-k-1&k+1&-2.}}

In section 5 we proved that the algebra generated by $L_k$'s and
$R_k$'s where $k\ge -1$ does not get renormalized. This is the
algebra for symmetric $SO(N)$ and antisymmetric $Sp(N)$ matter,
hence the algebra of chiral rotations of these gauge theories does
not receive nonperturbative corrections. The proof for the adjoint
representation works exactly as before if we substitute for all
subscripts of the generators in the equations of section 5 twice
their value. The proof of the nonrenormalization of the ${\cal
R}_k$ anomaly also carries over because the only difference in the
anomaly compared to the $U(N)$ gauge theory is the $c_k r_k$ term
in ${\cal L}_k$ which has the same form as $u_0 r_k$ so it cannot
receive corrections. The proof for ${\cal L}_k$ follows the same
pattern but instead of using the Wess-Zumino consistency condition
coming from $[L,R]$ commutator we use the condition coming from
$[L,L]$ commutator.

\newsec{Virasoro Constraints for the One-Matrix Model}

In this section we review the exact constraints for the planar
level free energy $F_m$ of the one-matrix  model \refs{\MM , \GM.}
We consider the $U(N)$ matrix model that is related to the $U(N)$
gauge theory with the adjoint scalar. The $SO(N)$ and $Sp(N)$
matrix models are treated similarly. We derive the loop equations
by considering the Virasoro algebra of redefinitions of the matrix
$M.$ This highlights the similarity of the algebraic structure of
the loop equations with the gauge theory anomalies. The partition
function of the matrix model is \eqn\mfree{Z_m=\exp \left( -{{\hat
N}^2 \over g_m^2}F_m \right)= \int d^{{\hat N}^2} M \exp \left(-
{{\hat N} \over g_m}  W(M) \right), } where $W(M)=
\sum_{i=1}^{n+1} {g_i\over i} \Tr \, M^{i}$ is the potential of
the matrix model and $F_m$ is the matrix model free energy. The
partition function is invariant under arbitrary redefinition of
the integration variable $ M \rightarrow f(M).$ These
redefinitions are symmetries of the matrix model. The generators
of the redefinitions annihilate the partition function and the
free energy \eqn\rmk{R_{m,k}=M^{k+1}{\delta \over \delta M}.} They
form a partial Virasoro algebra
\eqn\viram{[R_{m,k},R_{m,l}]=(l-k)R_{m,k+l},} where $k,l\geq -1.$
Acting  with $\epsilon R_{m,k}$ on the free energy $F_m$ we obtain
the following identity \eqn\inva{\eqalign{0&=\epsilon R_{m,k} F_m
\equiv  \epsilon {\cal R}_{m,k} \cr &= -{g_m^2 \over {\hat N}^2
Z_m} \delta \int d(M+\epsilon M^{n+1}) \exp \left( -{{\hat N}
\over g_m} \sum_{i=1}^{n+1} {g_i \over i} \Tr (M+ \epsilon
M^{k+1})^i \right).}} Expanding \inva\ to first order in
$\epsilon$ we have \eqn\invan{{\cal R}_{m,k}={-g_m^2 \over {\hat
N}^2 Z_m} \int d M \left(-{{\hat N} \over g_m} \sum_{i=1}^{n+1}
g_i \Tr \, M^{i+k}+ \Tr \, {\delta M^{k+1} \over \delta M} \right)
\exp \left(-{{\hat N} \over g_m} W(M) \right) .} To evaluate the
Jacobian we write \eqn\jeval{\eqalign{ \Tr \, { \delta M^{n+1}
\over \delta M}&= {\delta M^{k+1}_{ij} \over \delta M_{ij}} =
\sum_{i=l}^k { (M^l \delta M M^{k-l})_{ij}\over \delta M_{ij} }\cr
&= \sum_{l=0}^k M^l_{il} {\delta M_{lm} \over \delta
M_{ij}}M^{k-l}_{mj} = \sum_{l=0}^k \Tr \, M^l \Tr \, M^{k-l}.}}
Hence the variation of the free energy is \eqn\lkr{{\cal
R}_{m,k}=R_{m,k} F_m=\sum_{i=1}^{n+1} g_i \langle \Tr \,
M^{i+k}\rangle - \sum_{i=0}^k \langle \Tr \, M^i \Tr \,
M^{k-i}\rangle.} In the large  ${\hat N}$ limit the expectation
values of products $U(N)$ invariant operators factorize $\langle
\Tr \, M^i \Tr \, M^{k-i} \rangle = \langle \Tr \, M^i \rangle
\langle \Tr \, M^{k-i} \rangle.$ Defining $r_{m,k}= {g_m \over
{\hat N} } \langle \Tr \, M^k \rangle$  we rewrite \lkr\ in the
large ${\hat N}$ limit as \eqn\lkrn{{\cal
R}_{m,k}=\sum_{i=1}^{n+1} g_i r_{m,i+k} - \sum_{i=0}^k r_{m,i}
r_{m,k-i}} which takes the same form as the as the Konishi anomaly
\ln.\ The loop equations are obtained by setting ${\cal
R}_{m,k}=0.$ They are recursion relations for $r_{m,k}$ in terms
of the first $n$ moments $r_{m,0},\dots, r_{m,n-1}.$ Equivalently,
the loop equations determine the matrix model curve
$y^2(z)=W'^2(z)+f(z)$ where $y(z)={g_m \over {\hat N}} \langle \Tr
\, {1\over z-M} \rangle $ is the resolvent. The consistency
conditions for ${\cal R}_{m,k}$ are derived the same way as for
the gauge theory \wzr\ \eqn\matrixwz{R_{m,k}{\cal
R}_{m,l}-R_{m,l}{\cal R}_{m,k}=(l-k){\cal R}_{m,k+l}.} It is easy
to verify that \lkrn\ satisfies the consistency conditions
\matrixwz.\ Similarly one can show that the full matrix model loop
equations \lkr\ satisfy \matrixwz.\ The answer \lkrn\ is exact in
the planar limit of the matrix model, hence we do not need the
consistency conditions in this paper.

\newsec{Implications for the Dijkgraaf-Vafa conjecture}

Let us discuss the implications of the above results for the
relation between the matrix models and the supersymmetric gauge
theories. We will consider the $U(N)$ gauge theory with adjoint
matter to keep the discussion concrete. The anomalous variation of
the free energy of the gauge theory under $R_k$  \ln\ has the same
form as the variation of the matrix model free energy under
$R_{m,k}$ \rmk\ if we identify the expectation values \CachazoRY\
\eqn\identifyr{r_k=r_{m,k}.} The equations \ln\  ${\tilde R}_k=0$
can be considered as recursion relations for higher moments
${\tilde r}_i$ in terms of the first $n$ moments ${\tilde r}_0,
{\tilde r}_1 \dots {\tilde r}_{n-1}.$ Hence it is enough to
identify the first $n$ moments in \identifyr.\ The matrix model
then determines the expectation values of all chiral operators
$r_i.$ The expectation values of the moments of the scalar depend
also on the gauge symmetry breaking pattern $U(N)\rightarrow
\otimes_{i=1}^r U(N_i)$ \Cachazothree.\ The $U(1)$ photinos of the
$U(N_i)$ subgroups can have arbitrary vacuum expectation value.
These values determine all moments of the gaugino field $\Tr \,
\Phi^k W_\alpha$ \Svrcek.\ Hence the isolated massive vacua come
with a $2r$-dimensional fermionic moduli space where $r$ is the
rank of the low energy gauge group. In conclusion, matrix model
determines the expectation values of all chiral operators up to
the choice of the gauge symmetry breaking pattern and $k$
independent expectation values of the $U(1)$ photino condensates.

The generalized Konishi anomaly can be viewed as the equation of
the curve \eqn\mmcurve{y^2=W'^2(z)+f(z)} where $y$ is the
generating function of the glueball moments \CachazoRY.\ This
curve is identified with the matrix model curve  using \identifyr\
which is the same as identifying the polynomials $f(z)=f_m(z).$
The results from section 6 on nonperturbative corrections to the
Konishi anomaly imply that the gauge theory curve does not have
nonperturbative deformations for superpotentials of degree less
than $2N+1.$ Hence for these superpotentials the curve of the full
gauge theory agrees with matrix model curve. For higher degree of
the superpotential the curve can get deformed. We have identified
that the only possible deformation of the curve is the
nonperturbative renormalization of the superpotential. This is so
essentially because the form of the curve is uniquely fixed from
the Virasoro symmetry and we know from section 5 that the extended
Virasoro symmetry is exact in the full gauge theory. For given
$f(z)=f_m(z),$ the coefficients of the superpotential are the only
parameters of the curve.

The effective superpotential and the matrix model free energy are
generating functions for chiral operators and for the moments of
$M$ respectively \eqn\generate{\eqalign{ {\partial \over \partial
g_k} W_{eff}= \left\langle {\Tr \, \Phi^{k} \over k} \right\rangle
,\cr {\partial F_m \over \partial  g_k}= \left\langle {\Tr \, M^k
\over k}\right\rangle .}} To relate $W_{eff}$ and $F_m,$  we use
shift symmetry to generalize the first equation in \generate\ to a
generating function for $\Tr {\tilde W}^2 \Phi^k.$ The effective
superpotential is invariant under shift symmetry so it can be
written as \eqn\weffs{W_{eff}=\int d^2 \psi \, {\cal F}({\tilde
r}_i) } for some function ${\cal F}.$ We use \weffs\ to rewrite
the first equation in \generate\ as \eqn\fg{{\partial \over
\partial g_k} {\cal F}= \langle {{\tilde r}_k \over k} \rangle .}
Hence we have the relation \CachazoRY\
\eqn\super{F_m(S_i,g_k)={\cal F}({\tilde S}_i,g_k)|_{\psi=0}+
{\cal H}({\tilde S}_i)|_{\psi=0}} where ${\cal H}({\tilde S}_i)$
is a coupling independent function. Similar relations for the
$Sp(N)$ and $SO(N)$ gauge theory are given in \kraus\ and
\vafasp.\ The derivation of the relation \super\ rests on the
Konishi anomaly equations and on the validity of low energy
description of the gauge theory in terms of the glueball fields
$S_i.$ The nonrenormalization of the Konishi anomaly implies that
${\cal F}$ does not have additional nonperturbative corrections,
whence the relation \super\ is valid nonperturbatively. The
derivation of the nonperturbative exactness of the Konishi anomaly
is the first step in a full proof of the Dijkgraaf-Vafa
correspondence.

\bigskip

\centerline{\bf Acknowledgements}

I am very grateful to F. Cachazo, G. Ferretti, N. Seiberg and
especially to E. Witten for useful discussions/correspondence.
This research is supported in part by NSF grants PHY-9802484 and
PHY-0243680. Any opinions, findings and conclusions or
recommendations expressed in this material are those of the
authors and do not necessarily reflect the views of the National
Science Foundation.

\appendix{A}{Second Proof of the Nonrenormalization of the Algebra of Chiral Rotations}

In this appendix we give a proof of absence of nonperturbative
corrections to the extended Virasoro algebra without using the
shift symmetry. This proof is applicable to $SO(N)$ and $Sp(N)$
gauge theories which do not posses shift symmetry. We assume from
section 5 the  nonrenormalization of the Virasoro subalgebra
generated by $L_n$'s because we did not use shift symmetry to
prove it. We use the nonperturbatively defined Virasoro generators
$L_n$ to fix the nonperturbative definition of the remaining
generators by recursively commuting  $Q_{n,\alpha}$ and $R_n$ with
the raising operator $L_1.$ Having defined the generators, let us
show that the nonperturbative corrections to the $[L,Q]$
commutator vanish \eqn\lqcorrections{[L_m,
Q_{n,\alpha}]=(n-m)Q_{m+n,\alpha}+ \sum_{j=1}^{\infty}
\Lambda^{2jN} c^j_{m,n} Q_{m+n-2jN,\alpha}.} Firstly, we prove
nonrenormalization of $[L_0, Q_{n,\alpha}]$ using mathematical
induction. The lowest dimensional correction to the commutators is
$\Lambda^{2N}Q_{-1,\alpha}$ hence the first step of induction is
valid because the commutator of $L_0$ with $Q_{-1,\alpha}, \dots,
Q_{2N-2,\alpha}$ does not have nonperturbative corrections.
Assuming the induction hypothesis is valid for $Q_{-1,\alpha},
\dots, Q_{n,\alpha}$ we calculate \eqn\indstep{\eqalign{[L_0,
Q_{n+1,\alpha}]&={1\over n-1} [L_0,[L_1,Q_{n,\alpha}]]\cr
&={1\over n-1}[[L_0,L_1],Q_{n,\alpha}]+ {1\over n-1} [L_1,[L_0,
Q_{n,\alpha}]] \cr &=(n+1){[L_1,Q_{n,\alpha}] \over
n-1}=(n+1)Q_{n+1,\alpha},}} where the first equality comes from
the recursive definition of $Q_{n+1,\alpha},$ the second from
Jacobi identity, the third from the induction hypothesis and the
nonrenormalization of the Virasoro algebra and the last equality
is again from the recursive definition of $Q_{n+1,\alpha}.$ We
show the absence of corrections to the remaining $[L,Q]$
commutators by commuting them with $L_0$ and then using Jacobi
identity and the commutators we showed above to be nonrenormalized
\eqn\ljacobi{[L_0,[L_m,Q_{n,\alpha}]]=[[L_0,L_m],Q_{n,\alpha}]+
[L_m,[L_0,Q_{n,\alpha}]]=(m+n)[L_m, Q_{n,\alpha}].} But the
$[L_m,Q_{n,\alpha}]$ commutator is a linear combination of
$Q_{k,\alpha}$'s which are eigenvectors of the adjoint action of
$L_0$ with eigenvalue $k,$ whence the commutator is proportional
to $Q_{m+n,\alpha}$ so all corrections to the commutator vanish.
Let us show the absence of corrections to the  $[L,R]$ commutator
\eqn\lrcorrections{[L_m,R_n]=(n-m)R_{m+n}+\sum_{j=1}^{\infty}
\Lambda^{2jN} c^j_{m,n} R_{m+n-2jN}+ \sum_{j=1}^{\infty}
\sum_{i=1}^{n+1} \Lambda^{2jN}g_i d^{i,j}_{m,n} L_{m+n+i-2jN}.} We
commute \lrcorrections\ with $Q_{l,\alpha}$ to get
\eqn\qlkill{\eqalign{[Q_{l,\alpha},[L_m,R_n]]+[L_m,[R_n,Q_{l,\alpha}]]+[R_n,[Q_{l,\alpha},L_m]]&=\cr
=[Q_{l,\alpha}, \sum_{j=1}^{\infty} \sum_{i=1}^{n+1}
\Lambda^{2jN}g_i d^{i,j}_{m,n} L_{m+n+i-2jN}]&=\cr
=\sum_{j=1}^{\infty} \sum_{i=1}^{n+1}\Lambda^{2jN}g_i
(m+n-l+i-2jN) d^{i,j}_{m,n} Q_{m+n+i+l-2jN,\alpha}&=0.}} In
simplifying \qlkill\ we used the $[L,Q]$ commutator which we
proved above to be nonrenormalized and  the $[R,Q]=0$ commutator.
Clearly, the only way to satisfy the Jacobi identity \qlkill\ is
that $d^{i,j}_{m,n}=0.$ All $g_i$ dependent corrections vanish.
The remaining corrections have the same algebraic structure as the
corrections \qlkill\ to the $[L,Q]$ commutator so the
nonrenormalization proof for that commutator works for the $[L,R]$
commutator as well.

It remains to consider the $\{Q,Q\}$ anticommutator. The
nonperturbative corrections are proportional to $\epsilon_{\alpha
\beta}$  \eqn\qqcorrections{\eqalign{\{Q_{\alpha,m},Q_{\beta,n} \}
=& -\epsilon_{\alpha \beta} (n-m)R_{m+n} -\epsilon_{\alpha,\beta}
\sum_{j=1}^\infty \Lambda^{2jN} c^j_{m,n} R_{m+n-2jN}\cr
&-\epsilon_{\alpha,\beta} \sum_{j=1}^\infty \sum_{i=1}^{n+1}
\Lambda^{2jN}g_i d^{i,j}_{m,n} L_{m+n+i-2jN}.}} Consider the
following Jacobi identity
\eqn\lqqjacobi{\eqalign{0=[L_m,\{Q_{0,\alpha}, Q_{n,\beta} \}]
+\{Q_{0,\alpha},[Q_{n,\beta},L_m]\}-\{Q_{n,\beta},[L_m,Q_{0,\alpha}]\}&=
\cr -\epsilon_{\alpha,\beta} \sum_{j=1}^{\infty} \Lambda^{2jN}
R_{m+n-2jN}[(n-m-2jN)c^j_{0,n}+(m-n)c^j_{0,m+n}-m c^j_{n,m}]& \cr
-\epsilon_{\alpha,\beta} \sum_{j=1}^\infty \sum_{i=1}^{n+1}
\Lambda^{2jN}g_i L_{m+n+i-2jN}
[(n-m+i-2jN)d^{i,j}_{0,n}+(m-n)d^{i,j}_{0,m+n}-m d^{i,j}_{n,m}].}}
Setting $m=0$ we get $c^j_{0,n}=0$ and $d^{i,j}_{0,n}=0$ unless
$i=2jN.$ Substituting this back into \lqqjacobi\ we see that all
$c^j_{m,n}$ vanish and $d^{i,j}_{m,n}=0$ unless $i=2jN.$ To prove
that the remaining corrections vanish we evaluate the $R,Q,Q$
Jacobi identity
\eqn\rqqjacobi{\eqalign{[R_0,\{Q_{m,\alpha},Q_{n,\beta}\}]+\{Q_{m,\alpha},[Q_{n,\beta},R_0]\}
-\{Q_{n,\beta},[R_0,Q_{m,\alpha}]\}&= \cr
[R_0,\{Q_{m,\alpha},Q_{n,\beta}\}]&=\cr -\epsilon_{\alpha,\beta}
[R_0,\sum_{j>0} \Lambda^{2jN}g_{2jN}d^j_{m,n} L_{m+n}]=
-\epsilon_{\alpha,\beta} \sum_{j>0}
\Lambda^{2jN}g_{2jN}(m+n)d^j_{m,n}R_{m+n}&=0.}} Hence, $d^j_{m,n}
\equiv d^{2jN,j}_{m,n}=0$ and the $\{Q,Q\}$ anticommutator is
nonrenormalized.

\listrefs

\end